\documentclass[
    ,final            
  ]
  {aipproc}

\layoutstyle{8x11double}

\begin{document}

\title{On the Velocity Field and the 3D Structure of the \\ Galactic Soccer Ball Abell\,43}


\author{Thomas Rauch}{
  address={Institut f\"ur Astronomie und Astrophysik, Universit\"at T\"ubingen, Germany}
}

\author{Klaus Werner}{
  address={Institut f\"ur Astronomie und Astrophysik, Universit\"at T\"ubingen, Germany}
}
\author{Barbara Ercolano}{
  address={Department of Physics and Astronomy, UCL, United Kingdom}
}

\author{Joachim K\"oppen}{
  address={Observatoire de Strasbourg, France}
}

\begin{abstract}
Planetary nebulae (PNe) and their central stars (CSs) are ideal tools to test evolutionary theory:
photospheric properties of their exciting stars give stringent constraints for theoretical 
predictions of stellar evolution. The nebular abundances display the star's photosphere 
at the time of the nebula's ejection which allows to look back into the history of stellar evolution 
-- but, more importantly, they even provide a possibility to investigate on the chemical evolution of our 
Galaxy because most of the nuclear processed material goes back into the 
interstellar medium via PNe. 

The recent developments in observation techniques and a new three-dimensional 
photoionization code \mbox{MOCASSIN} (Ercolano et al\@. 2003) enable us to analyze PNe properties precisely
by the construction of consistent models of PNe and CSs. In addition to PNe imaging
and spectroscopy, detailed information about the velocity field within the PNe is
a pre-requisite to employ de-projection techniques in modeling the physical structure
of the PNe.
\end{abstract}

\maketitle

\begin{figure}
  \includegraphics[height=.19\textheight]{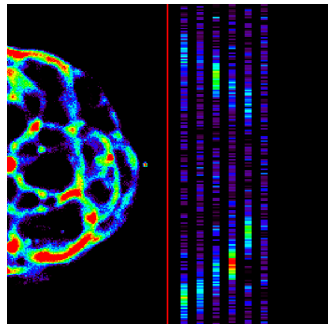}
  \caption{Left: \mbox{[O\,{\sc iii}]}~$\lambda\,5007$\,\AA\ image of the PN A\,43 (western part).
           Right: Intensity of \mbox{[O\,{\sc iii}]}~$\lambda\,5007$\,\AA\ in A\,43 measured in six apertures, 
           west of its central star.
           The horizontal axis shows $-50'' < \Delta\mathrm{RA} < +50''$. The
           vertical axis is the differential radial velocity in km/sec ($-55 < \Delta v < +55$).}
  \label{f:dvrad}
\end{figure}

In July 1998, we performed imaging and spectroscopy of the PN A\,43 and its exciting star at ESO, La Silla.
The H\,$\alpha$ and [O\,{\sc iii}]~$\lambda\,5007$\,\AA\ (Fig\@. \ref{f:dvrad})
images show prominent deviations from spherical symmetry which deserve further investigation.
Subjective image interpretations of different observers range
from ``radial filaments'' over ``soap bubbles'' to ``penta- and hexagons like a (soccer) football's seams''.
Also, instabilities in the nebula's surface are prominent. The most likely explanation might be
that the old, slow AGB wind matter is swept up to a thin shell by the fast central star wind. 
While the invisible inner, high-pressure bubble is expanding due to the released energy of the stellar wind,
instabilities in the dense, moving shell may appear (Vishniac 1983), effective enough to
produce filament-like surface structures of the shell matter. As these filaments form, the intrafilament
region can expand out ahead of the filaments, giving rise to a somewhat ``lumpy'' outer edge on the shell.
This is quite obvious on the image of A\,43. Similar PNe are known, e.g. NGC\,6894, NGC\,7048, or
NGC\,7139 (Balick 1987) but the edges of their shells appear smooth and round in projection.
Thus, A\,43 is an excellent test case also for hydrodynamical models!

The CES spectra of A\,43 show an expansion velocity of the shell, measured in 
\mbox{[O\,{\sc iii}]}~$\lambda\,5007$\,\AA\ of 
up to 50 km/sec (Fig\@. \ref{f:dvrad}). A\,43 has an almost spherical shell with strong density variations.
The spectra allow to construct a ``third dimension'', i.e., a 3D density distribution.
However, it turned out that our 12 aperture positions in the nebula are not sufficient to
provide a reliable database for the de-projection method. 
Since a reliable 3D density distribution is a crucial input for any
3D photoionisation code, a spatially more complete measurement 
(about ten times more positions) of the radial velocity is necessary.

\paragraph{Acknowledgments} This research was supported by the DLR under grants 50\,OR\,9705\,5 and 50\,OR\,0201.


\bibliographystyle{aipproc}   

\begin{thebibliography}{9}
\bibitem{B87}   Balick B\@. 1987, AJ 94, 971
\bibitem{EEA03} Ercolano B., Barlow, M.J., Storey, P.J., \& Liu X.-W\@. 2003, MNRAS 340, 1136 
\bibitem{V83}   Vishniac E.T\@. 1983, ApJ 274, 152
\end{thebibliography}

\end{document}